\documentclass[aps,preprint,superscriptaddress]{revtex4}
\usepackage{graphicx}

\newcommand {\CC}{$^{\circ}$C}   %degree C
\renewcommand{\thefootnote}{\fnsymbol{footnote}}

\begin{document}
%\preprint{}

\title{Spin-polarized electron injection through an Fe/InAs junction}

\author{Hiroshi Ohno}
\email{ohno@nano.eng.hokudai.ac.jp}
\affiliation{Nanoelectronics Laboratory, Graduate School of Engineering, Hokkaido University, Sapporo 060-8628, Japan.}
\author{K. Yoh}
\email{yoh@rciqe.hokudai.ac.jp}
\affiliation{Research Center for Integrated Quantum Electronics (RCIQE), Hokkaido University, Sapporo 060-8628, Japan.}
\author{K. Sueoka}
\affiliation{Nanoelectronics Laboratory, Graduate School of Engineering, Hokkaido University, Sapporo 060-8628, Japan.}
\author{K. Mukasa}
\affiliation{Nanoelectronics Laboratory, Graduate School of Engineering, Hokkaido University, Sapporo 060-8628, Japan.}
\author{A. Kawaharazuka}
\affiliation{Paul-Drude-Institut f\"{u}r Festk\"{o}rperelektronik, Hausvogteiplatz 5-7, 10117 Berlin, Germany.}
\author{M. Ramsteiner}
\affiliation{Paul-Drude-Institut f\"{u}r Festk\"{o}rperelektronik, Hausvogteiplatz 5-7, 10117 Berlin, Germany.}

\date{\today}

\begin{abstract}
We report on the spin-polarized electron injection through an Fe(100)/InAs(100) junction.
The circularly polarized electroluminescence of injected electrons from epitaxially grown Fe thin film into InAs(100) in an external magnetic field is measured to investigate the spin injection efficiency.
The obtained polarization of the electroluminescence is seen to increase up to about -12 \% at the temperature of 6.5 K and the external magnetic field of 10 T.
This result suggests that the efficient spin injection is possible through the ferromagnetic metal/semiconductor (FM/SC) interface without a tunneling barrier despite the contradictory arguments based on conductivity mismatch at the FM/SC interface.
\end{abstract}

%72.25.Dc Spin polarized transport in semiconductors 
%72.25.Hg Electrical injection of spin polarized carriers
%75.50.Bb Fe and its alloys
%78.60.Fi Electroluminescence

\maketitle

Spin-polarized electron injection (spin injection) into semiconductor is indispensable to realize spin-related devices, such as spin transistors.\cite{1}
Several candidates were proposed as a spin-polarized current source, e.g., ferromagnetic metals (FM),\cite{2,3,4,5} diluted magnetic semiconductors (DMS),\cite{6,7,8} and half-metallic ferromagnetic metals.\cite{9}
At first, the spin injection into semiconductors were demonstrated using DMS such as Be$_x$Mn$_y$Zn$_{1-x-y}$Se\cite{6} and (Ga,Mn)As.\cite{7,8}
However, the difficulty of room temperature operation of the device remained unsolved because of their low Curie temperatures.
In addition, the Be$_x$Mn$_y$Zn$_{1-x-y}$Se spin aligner required a large external magnetic field.
On the other hand, ferromagnetic metal/semiconductor (FM/SC) hybrid structures are supposed to be promising to realize high efficient spin injection even at room temperature and without any external magnetic field.
Despite great efforts, the spin injection in the FM/SC hybrid structure remained a challenging task for a long time.
Furthermore, there appeared theoretical predictions that spin injection in the FM/SC hybrid structure in the diffusive regime is hardly possible due to a large conductivity mismatch between a ferromagnetic metal and a semiconductor.\cite{10,11}
Recently, spin injections in FM/GaAs hybrid structures with a Schottky tunneling barrier were demonstrated.\cite{2,3,4,5}
Theoretically, the interfacial resistance, e.g., due to a tunneling barrier would be expected to enhance the spin injection efficiency.\cite{12,13,14}
However, it is not still clear that the tunneling barrier is indeed necessary for spin injection.
Yu \textit{et al.}\cite{15,16} pointed out that an electric-field enhances the spin injection efficiency in FM/SC hybrid structures even in the diffusive regime.
Recent theoretical calculations of spin injection in the ballistic regime predict highly efficient spin injection at the FM/SC interface.\cite{17,18,19}

In this paper, we demonstrate the spin injection through an Fe(100)/p-InAs(100) junction without a tunneling barrier. 
InAs has a significant advantage over GaAs for spin-related device applications, because of the following reasons: (1) InAs is expected to form an ideal ohmic contact to metal,\cite{20,21} (2) a significant spin-orbit interaction is expected in an InAs 2DEG channel\cite{22,23} compared with GaAs.
We have confirmed that the epitaxial Fe(100) thin film can be grown on the InAs(100) at the low growth temperature of 23 \CC, which is expected to reduce the inter-diffusion at the interface, and the Fe/n-InAs junction showed a good ohmic behavior.\cite{21}
Measurements of circularly polarized electroluminescence (EL) in Fe/p-InAs junctions were performed in order to investigate the electron injection and their spin states.
The spin polarization of injected electrons was estimated from the degree of circular polarization of the luminescent light.

\begin{figure}[p]
\includegraphics{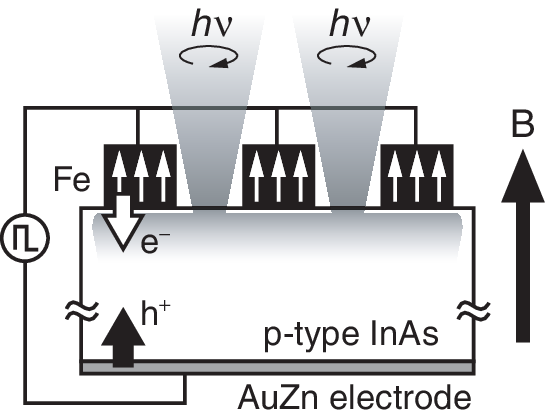}
\caption{The schematic sample structure of the Fe/p-InAs junction. The pitch and width of the Fe wire array is 4 $\mu$m and 2 $\mu$m, respectively. Only the luminescent light from gaps between Fe wires can be detected by the InSb infrared detector. The external magnetic field is applied to the direction perpendicular to the sample surface, which is the same direction for optical detection.}
\label{f1}
\end{figure}

The Fe/p-InAs junction for the EL measurements was prepared in the following procedure.
Prior to Fe deposition, 300nm-thick SiO$_2$ island was formed for the non-magnetic bonding pad formation in the final step of the device preparation.
In an ultra high vacuum (UHV) chamber, the InAs substrate was annealed at about 450 \CC\ without arsenic flux irradiation for thermal cleaning.
Fe was deposited on the InAs substrate by an electron-beam evaporator with the substrate temperature of 23 \CC.
After the Fe growth, the sample was taken out of the chamber to be subjected to the patterning process.
The Fe thin film was patterned into an array of wires (pitch; $p=$4 $\mu$m, width; $w=$2 $\mu$m) whose axes were taken along the $\langle011\rangle$ direction.
Finally, a Au/Ti electrode pad was fabricated on the SiO$_2$ island region.
The effective contact area of the sample is approximately 0.5$\times$3 mm$^2$.
The cross-sectional schematic sample structure of the Fe/p-InAs junction is shown in Fig.~\ref{f1}.
Note that the external magnetic field was applied perpendicular to the surface, and the luminescence was detected in the upper direction perpendicular to the surface and parallel to the external magnetic field.

The sample was mounted in a continuous flow He cryostat with a superconducting magnet and cooled down for the EL measurements.
The luminescence was detected from the top of the sample surface by an InSb infrared detector.
The right ($\sigma^+$) and left ($\sigma^-$) circularly polarized components of the luminescence are analyzed using a first-order quarter waveplate and a polarizer.
The degree of circular polarization of the luminescence $P_L$ is determined by $P_L=(\sigma^+-\sigma^-)/(\sigma^++\sigma^-)$.

The spin polarization $P_s$ of injected electrons can be estimated from the circular polarization $P_L$ of the luminescence due to the optical selection rule in zincblende compound semiconductors.\cite{24}
In a condition of constant carrier generation, such as constant bias voltage, $P_L$ is given by
\[
P_L=-\frac{1}{2}\frac{1}{(1+2\tau_{rec}/\tau_s)}P_s,
\]
where $\tau_{rec}$ and $\tau_s$ are a recombination time and a spin relaxation time of electrons, respectively.
Note that the polarization of the luminescence is reduced from the spin polarization of electrons by the factor $\tau_{rec}/\tau_s$.
However, limited report is known for the carrier dynamics and the spin dynamics in bulk InAs so far.
To the best of our knowledge, the spin relaxation in bulk InAs was reported only by Boggess {\em et al.}\cite{25}.
According to their result, $\tau_{rec}$ and $\tau_s$ at 300 K are $\sim$ 1 ps and $\sim$ 20 ps, respectively.
If we assume the relationship $\tau_{rec} \ll \tau_s$ to hold even at low temperatures, the factor of ${1}/{(1+2\tau_{rec}/\tau_s)}$ becomes almost negligible.
In that case, we obtain a simple factor of $-2$ (i.e. $P_s=-2P_L$).
In general, temperature dependence of the $P_L$ would be expected by the temperature dependence of these time constants.
It is to be added that the discussion above assumes that the heavy hole and light hole bands are degenerate.

The clear EL from the Fe/InAs hybrid structure was observed above the bias voltage of 300 mV.
It was verified that the peak of the EL spectrum was about 3.08 $\mu$m (i.e. 0.403 eV), suggesting that the luminescence is caused by transitions between the conduction band and the valence band/the acceptor level in the p-InAs substrate.
There was no threshold voltage difference in luminescence and current.
Present result implies that injected electrons into the conduction band of InAs readily recombine with holes in the InAs substrate in the vicinity of the interface.
The low ohmic resistance for electrons and the low bias operation of the Fe/InAs hybrid structure, which arise from the surface pinning level of InAs is located at 50 $\sim$ 100 meV above the conduction band,\cite{26,27} are great advantages for electron injection into semiconductors.
The detected luminescence is likely to come from the gap of the Fe wires because no luminescence was observed in a control sample which is completely covered with the Fe film.

\begin{figure}[p]
\includegraphics{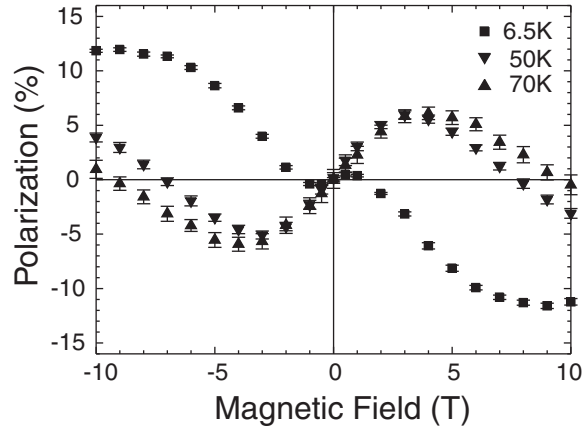}
\caption{The external magnetic field dependence of the circularly polarization of the EL luminescence from the Fe/p-InAs heterostructure at T = 6.5, 50 and 70 K.}
\label{f2}
\end{figure}

\begin{figure}[p]
\includegraphics{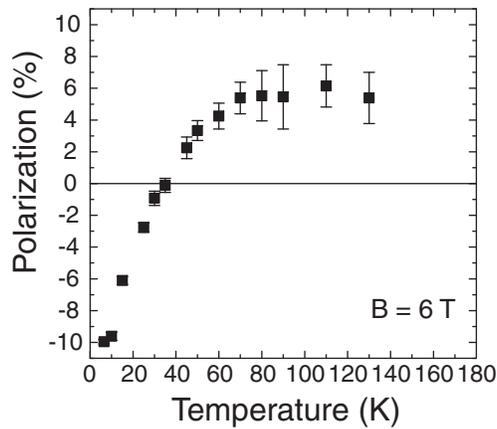}
\caption{The temperature dependence of the polarization of the EL luminescence at B = 6 T.}
\label{f3}
\end{figure}

For polarization measurements, the $\sigma^+$ and $\sigma^-$ components of the luminescence were obtained by switching the angle of the quarter waveplate.
Figure~\ref{f2} shows the external magnetic field dependence of the $P_L$ from the Fe/p-InAs hybrid structure at T = 6.5, 50 and 70 K.
The polarization was observed to increase up to about -12 \% at B = 10 T. 
However, below B = 1 T, the polarization showed clear sign flip by $\sim$ 0.5 \%.
For higher temperatures (T = 50 or 70 K), the polarization showed obviously the positive value of about 7 \% at B = 4 T.
The temperature dependence of the $P_L$ at B = 6 T is shown in Fig.~\ref{f3}.
As the temperature increases, the polarization decreases rapidly, and the polarity of the polarization is reversed from negative to positive.
At about T = 80 K, the polarization seems to saturate or begin to reduce again.
These complicated behaviors seem to be caused by two competing mechanisms.
The positive dependence is probably due to the large Zeeman splitting in the conduction band of InAs.\cite{28}
The negative dependence behavior, on the other hand, probably arises from the increase of majority spins in the Fe spin-injector as the magnetization proceeds.
It should be mentioned that we have obtained preliminary results of non-magnetic sample showing similar behavior as high temperature (70 K) results.

Although the out-of-plane magnetization of Fe films are usually saturated at about B = 2 T, the measured $P_L$ does not follow the magnetization curve unlike the case of the Fe/GaAs systems.\cite{2}
In the case of the Fe/InAs systems, the polarization of injected electrons is influenced by both the spin polarization of Fe and the spin splitting of the conduction band in InAs induced by the Zeeman effect.
In addition to this, the heavy hole and the light hole valence bands also split by the external magnetic field.
Therefore the $P_L$ would not necessarily follow the magnetization curve and show more complicated behavior.
In addition to the spin polarization in Fe and the Zeeman spin splitting in InAs described above, there may be other effects which yield the circular polarization such as the magneto-optical effect of the Fe films and the effect of the fringe magnetic field around the Fe wires.
The magneto-optical effect can be excluded in our experimental set-up (Fig~\ref{f1}) because the luminescent light barely penetrates through an Fe thin film.
The fringe magnetic field is antiparallel to the external magnetic field near the center of the gap where most of the measurable recombination takes place, so that the total magnetic field in this region is always smaller than the external magnetic field.\cite{29}
Furthermore, the effect of fringe magnetic field, which exhibits a weak temperature dependence in this temperature range, can not account for the temperature dependence of the $P_L$ shown in Fig.~\ref{f3}.
Therefore we conclude that the $P_L$ data shown in Fig.~\ref{f2} reflect the spin polarization of Fe spin-injector as well as the Zeeman effect in the InAs.
Please note that the Zeeman effect acts oppositely against majority spin injection effect on $P_L$.
The temperature dependence of the $P_L$ in the range between 6.5 K and 80 K shown in Fig~\ref{f3} indicates that the negative dependence gradually disappears with increasing temperature.
This is presumably caused by the decrease of $\tau_{s}$ and subsequent increase of $\tau_{rec}/\tau_{s}$ with increasing temperature

We conclude that the injection of majority spins dominates in the present Fe/InAs hybrid structure by taking into account the polarity of the polarization.
This is in good agreement with the theoretical result calculated from first-principles by Zwierzycki \textit{et al}.\cite{19}, but contrary to the spin injection in the FM/GaAs  hybrid structure with the tunneling barrier, in which the minority spin makes dominant contribution.\cite{2,3,5}

In conclusion, we have demonstrated that the spin-polarized electron injection through the Fe(100)/p-InAs(100) junction.
The circular polarization of about -12 \% at T =6.5 K and B = 10 T, implies that spin-polarized electrons are successfully injected from Fe into InAs, which reflect the spin polarization of Fe, even though the large Zeeman splitting in InAs also influences the polarization of the luminescence.
This result is encouraging for future applications of FM/InAs hybrid structures or InAs-based heterostructures to spin-related devices.

We would like to thank to Prof. K. H. Ploog of Paul-Drude-Institut f\"{u}r Festk\"{o}rperelektronik for helpful discussions.
This work is partly supported by Grant-in-Aid for Scientific Research from the Japanese Ministry of Education, Culture, Sports, Science and Technology.

\printfigures

\end{document}